\documentclass[epj,final]{svjour}
\usepackage{epsf,here,latexsym}
\usepackage{graphics}

\begin{document}
\title{Some aspects of electrical conduction in granular systems of various dimensions}

\author{M. Creyssels\inst{1} \and S. Dorbolo\inst{2} \and A. Merlen\inst{1} \and C. Laroche\inst{1} \and B. Castaing\inst{1} \and E. Falcon\inst{1}\fnmsep\thanks{Present address: Mati\`ere et Syst\`emes Complexes, Universit\'e Paris-Diderot -- Paris 7, CNRS -- 75 013 Paris, France}}
\institute{Laboratoire de Physique, Ecole Normale Sup\'erieure de Lyon, CNRS -- 46 all\'ee d'Italie, 69 007 Lyon, France \and  GRASP Photop\^ole, Physics Department, Universit\'e de Li\`ege, B-4000 Li\`ege, Belgium} 
\mail{eric.falcon@univ-paris-diderot.fr}
\authorrunning{M. Creyssels  {\em et al.}}


\abstract{We report on measurements of the electrical conductivity in both a 2D triangular lattice of metallic beads and in a chain of beads. The voltage/current characteristics are qualitatively similar in both experiments. At low applied current, the voltage is found to increase logarithmically in a good agreement with a model of widely distributed resistances in series. At high enough current, the voltage saturates due to the local welding of microcontacts between beads. The frequency dependence of the saturation voltage gives an estimate of the size of these welded microcontacts. The DC value of the saturation voltage ($\simeq 0.4$ V per contact) gives an indirect measure of the number of welded contact carrying the current within the 2D lattice. Also, a new measurement technique provides a map of the current paths within the 2D lattice of beads. For an isotropic compression of the 2D granular medium, the current paths are localized in few discrete linear paths. This quasi-onedimensional nature of the electrical conductivity thus explains the similarity between the characteristics in the 1D and 2D systems. 
}
\maketitle
\section{Introduction}
\label{intro}
Granular materials are ubiquitous in the nature and have been extensively studied mostly since the 1990s \cite{PG05}. However, transport phenomena within granular media such as sound propagation \cite{Acous,Gilles03}, heat transfer \cite{Batchelor77,Chen06} and electrical conduction \cite{Holm00}  are not so much addressed. 

Electrical conduction within metallic granular media is a very old problem.  In 1890 E. Branly discovered the extreme sensitivity of the conductivity of metal fillings to an electromagnetic wave \cite{Branly90}. Although this effect was involved in the first wireless radio transmission near 1900, it is still not well understood (see Ref.\ \cite{Falcon05} for a review).  Except some sporadic works in the 1960-70s \cite{Gabillard66,Kamarinos75}, the electrical conduction in granular media prompts renewed interest only since recently.  Some works revisited the action of nearby electromagnetic wave (Branly effect) \cite{Dorbolo03,Dorbolo07}, others focused on the effect of an electrical source directly imposed on the system \cite{Falcon05,Vandembroucq97,Falcon04v1D}. Both perturbations lead to a transition from an insulating state to a conductive state of the granular medium. Using a 1D model granular system, it has been shown that the transition of conduction is related to the local welding of the microcontacts between grains when the applied voltage reaches a critical value \cite{Falcon05,Falcon04v1D}. 

Electrical conduction within granular media also display other astonishing properties: stochastic current fluctuations \cite{Kamarinos75,Bonamy00,Falcon04v3D,Creyssels06}, slow relaxations \cite{Creyssels06}, percolation \cite{Ambegaokar,Bardhan97}. Some of these electrical properties of metallic granular packings can be due to the extreme sensitivity of the intergrain electrical resistance to the nature of the grain surface \cite{DaCosta00}. One could even suspect completely irreproducible behaviours, fully dependent on the history and minute details of the material. Curiously enough, generic and reproducible behaviours can be observed in some circumstances. 

This paper attempts to describe some of these generic behaviours. To wit, two model granular systems will be compared: a 1D linear chain and a 2D lattice of beads under an applied stress. Surprisingly, the voltage/current characteristics are qualitatively similar in both experiments. Here, we also report the first experimental map showing the current distribution within a 2D granular medium. Note that direct infrared visualizations and numerical simulations of temperature distribution exist \cite{Vandembroucq97,Zhang02}, as well as various 2D models of random resistor network \cite{Vojta96}. This work is a first step towards more complex systems, some questions being still open, {\it e.g.}, the stress dependence of the electrical resistance of a granular packing \cite{Bowden39}.

\section{Experimental setup}
\label{exp}
 
\begin{figure}[t!]
\resizebox{1\columnwidth}{!}{%
\includegraphics{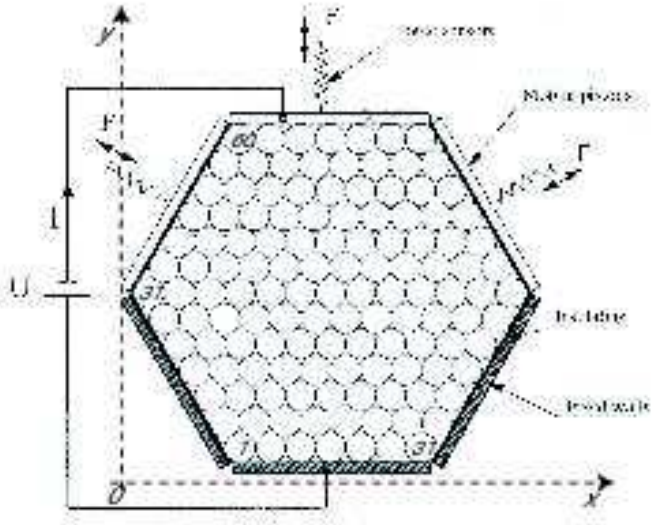}}
\caption{Sketch of the 2D experimental setup (top view). The real size of each side of the hexagonal lattice is 31 beads.}
\label{fig01}
\end{figure}

 The 2D experimental setup is sketched in Fig.\ \ref{fig01}. Stainless steel beads, 8 mm in diameter (with a $\pm 4\mu$m tolerance~\cite{Marteau}), are confined in a hexagonal cell between two horizontal plates of Teflon (PTFE) in order to achieve low friction between beads and framework. The beads are placed on a triangular lattice. The total number of beads is 2792, and the size of the lattice side is 31 beads. Three sides of the hexagon are fixed, while the others may move independently along their normal direction. As shown in Fig.\ \ref{fig01}, an horizontal stress can be applied either in one single direction or isotropically with the help of a feedback loop \cite{Gilles03}. The isotropic compression is created by compressing in the three directions by means of 3 linear motors, whereas the compression in one single direction is created with one motor, the two other ones being kept fixed. The maximum force applied on each side is limited to 80 N. The applied force on each lattice side is measured by three static force sensors. In the following, only an isotropic stress is applied. Each side of the hexagon is electrically insulated, except both faces perpendicular to the $y$-axis (see Fig.\ \ref{fig01}). A DC current source is supplied to both electrodes by a source meter (Keithley 2400) which also gives a measurement of the voltage drop. The current is injected to the lattice side and not to only one bead. During a typical run, a current $I$ is applied in the range $10^{-6} \leq I \leq 0.1$ A, and the voltage $U$ (or the resistance $R\equiv U/I$) is simultaneously measured. The current is supplied during a time shorter than 1 s in order to avoid possible DC Joule heating. Similar results are found in experiments conducted by imposing the voltage ($10^{-2} \leq U \leq 200$ V) and measuring the current (or the resistance). 

In this 2D experiment, the map of current paths is monitored by means of a magnetic field sensor (Bartington Inst. Mag-03) \cite{CapteurFlux}. Current flow through a conductor generates a magnetic field $B$ perpendicular to the axis of the conductor ({\OE}rsted experiment) and proportional to the current strength (Biot and Savart law). The sensor sensitivity is of the order of 500 $\mu$T. To avoid the interference from the Earth's magnetic field, a high frequency ($\approx$ 10 kHz) current is applied to the bead lattice by means of the source of a lock-in amplifier.  A typical current map is obtained as follows. The thick Teflon lid (40 mm in thickness) is replaced by a thin hexagonal glass plate (8 mm in thickness) in order to achieve the magnetic measurement. A small stress $\sim 1$ N is imposed on each side to settle the contacts. The glass plate is then tightened on the beads network by means of 6 screws and a bar (parallel to the $x$-axis) to avoid a possible buckling of the network at larger force. We first apply an isotropic stress on the 2D network ($\leq 80$ N by side). The system is then allowed to relax for one hour in order to avoid possible mechanical drift. A voltage is then applied to both electrodes parallel to the $x$-axis (see Fig.\ \ref{fig01}), the total current being measured with an ammeter (Keithley).  We wait until the current is stabilized. The magnetic sensor then is placed above the glass plate and is parallelly moved along the $x$-axis (orthogonally to the current direction) into grooves of the plate, from bead to bead along each of the 60 lines of beads. In the vicinity of each bead, the value of $B$ is then measured with the lock-in amplifier. We have checked that this value is proportional to the current averaged within a radius of two adjacent beads. One hour is necessary to perform the whole current map.  

The one-dimensional experiment consists of a chain of stainless steel beads, 8 mm in diameter (with a $\pm 4$ $\mu$m tolerance~\cite{Marteau}), submitted to a fixed stress. The number of beads, $N$, in the chain can be varied from $N=2$ to 40. A current, $I$, is injected to the chain, and the voltage $U$ of the whole chain is simultaneously measured (in the same way as in the 2D experiment) as well as the voltage, $U_i$, between each beads by means of multimeter/switch system (Keithley 2700 with multiplexer module 7702). The 1D experimental setup has been already described elsewhere~\cite{Falcon04v1D}.

Note that for both experiments, after each cycle of current, new contacts between each bead are needed to obtain reproducible measurements for the next cycle \cite{Falcon04v1D}. To wit, the applied force is reduced to zero, and we roll the beads which creates new contacts for the next cycle.

\section{Voltage/current characteristics: saturation voltage}
\label{characteristics}

\begin{figure}[t!]
\resizebox{1\columnwidth}{!}{%
\includegraphics{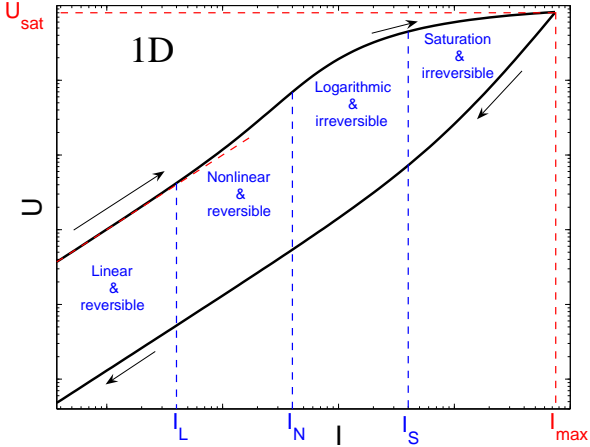}}
\caption{(color online) Log-Log schematic view of the $U$ -- $I$ characteristic for a 1D granular medium. The currents $I_L$, $I_N$, and $I_S$ define the limits of the different regimes of the up-characteristic: linear; nonlinear (see Sect.~\ref{NL}), logarithmic (see Sect.~\ref{log}) and saturating (see Sect.~\ref{characteristics} and \ref{freq})}
\label{schem}
\end{figure}

The typical electrical characteristic of a 1D chain of beads is schematically displayed in Fig.\ \ref{schem} which can be summarized as follows \cite{Falcon05,Falcon04v1D}. For a given applied current $I$, the measured $U(I)$ depends on the history of the imposed current.  Whenever the current remained below the value $I$ since the last renewal of contacts, the behaviour of the contact is designated as ``up-characteristic''.  On the other hand, once a larger current has been injected, the behaviour is ``down-characteristic''.
Down-characteristics are always reversible: as long as $|I|<I_{max}$  ($I_{max}$ being the maximum current previously applied), the relation  $U(I)$ is symetric [$U(I)=-U(-I)$] and remains the same whatever the history of $I$ since the last occurence of $I=I_{max}$.

Depending of the applied force, three typical currents $I_L$, $I_N$, and $I_S$ can be defined (see Fig. \ref{schem}):

\begin{itemize}

\item If $I_{max}<I_L$, up and down-characteristics are identical and linear  (see Fig.\  \ref{schem}). 

\item If $I_L<I_{max}<I_N$, up and down-characteristics remain identical but are nonlinear (see Fig.\ \ref{schem}). Section \ref{NL} is devoted especially to this regime. 

\item For $I>I_S$, the up-characteristic shows a constant voltage $U_{sat}$ (see Fig.\ \ref{schem}) proportionnal to the number of contacts  $\simeq 0.4$ V per contact. The down-characteristic is nonlinear, revealing the existence of local welding of microcontacts between beads \cite{Falcon05,Falcon04v1D}: the crowding of the current lines within these microcontacts generates a thermal gradient in their vicinity when significant Joule heat is produced. At high enough current ({\it i.e.}, when $U_{sat}$ is reached), this process leads to the local welding of the microcontacts. The contact resistance increases with the temperature of the welded microcontact, which itself only depends on the voltage across it. This regime will be discussed in the present section. 

\item Finally, for $I_N<I_{max}<I_S$, up-characteristic progressively raises up to the saturation voltage (see Fig.\ \ref{schem}). Down-characteristics are intermediate between those of the two neighbouring ranges. This will be discussed in Sec.\ \ref{log}.

\end{itemize}
\begin{figure}[t!]
\begin{tabular}{c}
\resizebox{1\columnwidth}{!}{%
\includegraphics{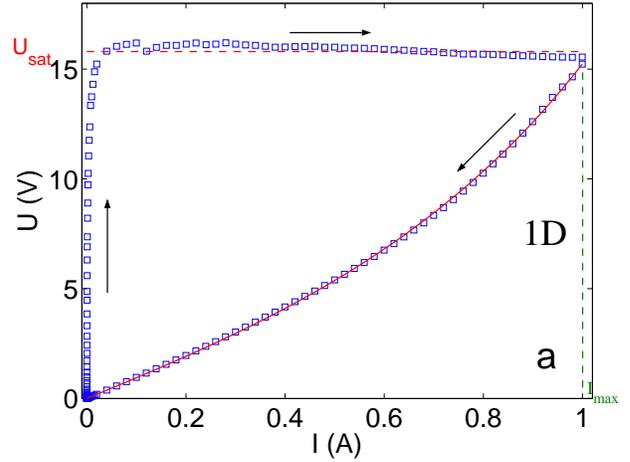}
} \\
\resizebox{1\columnwidth}{!}{%
 \includegraphics{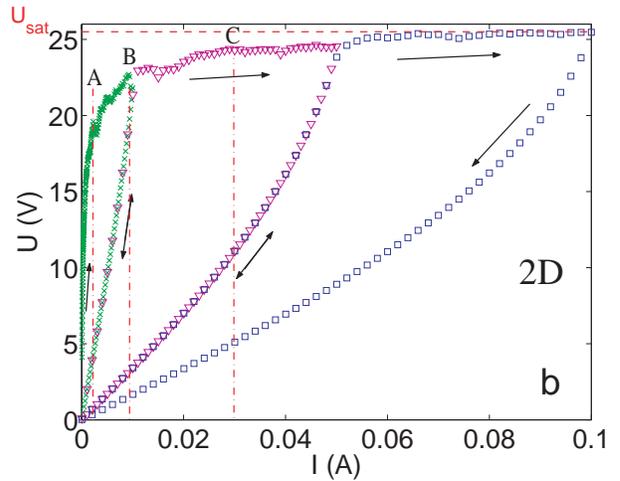}
}
\end{tabular}
\caption{(color online) Typical $U$ -- $I$  characteristics. {\bf (a)} 1D chain of $N=38$ beads for an applied force $F=100$ N. The dashed line $U_{sat}\simeq 15.8$ V corresponds to a saturation voltage per contact of $V_o\equiv U_{sat}/(N+1)\simeq 0.4$ V. The solid line is the theoretical down trajectory (see Ref. \cite{Falcon05,Falcon04v1D}). {\bf (b)} 2D lattice of beads for an applied force $F=50$ N. The dashed line corresponds to the saturation voltage $U_{sat} \simeq 25.4$ V. This value gives a measurement of the number of welded contact (see text).  Different symbols correspond to different maximum applied current, $I_{max}$. The letters A, B and C correspond to the current maps of Figs.\ \ref{fig03}a, b and c.}
\label{fig02}  
\end{figure}
The new results about the saturation range concern two points: the extension to 2D systems (see below) and the extension to AC voltages for the 1D system (see Sec.\ \ref{freq}).

The $U$ -- $I$ characteristic of a 1D chain of $N=38$ beads is shown in Fig.\ \ref{fig02}a  ($N$ being the number of beads between the two bead electrodes). Such typical curve has already been discussed in Ref.\ \cite{Falcon05,Falcon04v1D}. Notably, a saturation voltage per contact, $V_o\equiv U_{sat}/(N+1)\simeq 0.4$ V, is observed due to the local welding of microcontacts between beads \cite{Falcon05,Falcon04v1D}. Figure\ \ref{fig02}b displays the electrical characteristic for the 2D lattice of beads. Qualitatively, it is the same characteristic than the 1D one. This is quite surprisingly at first sight, but this can be understood when looking at the map of the  current paths in the 2D system: the current appears localized in few discrete linear paths whatever the strength of the injected current (see Fig.\ \ref{fig03} and Sec.\ \ref{map} below). Such a concentration of current on chains thus explains the similarity between the electrical characteristics in the 1D and 2D lattices. Moreover, for the 2D lattice, the 25.4 V saturation voltage (see Fig.\ \ref{fig02}b) gives a number of $64 \pm 1$ welded contacts when compared to the $V_o\equiv 0.4 \pm 0.01$V per contact of the 1D chain. This estimate is in good agreement with the dimensions of the hexagon (31 beads per side). Thus, as in the 1D system, the saturation voltage is also a measurement of the number of welded contact in the 2D system.

\section{Visualization of current paths in a 2D granular medium}
\label{map}
\begin{figure}[t!]
\begin{tabular}{c}
\resizebox{.98\columnwidth}{!}{%
\includegraphics{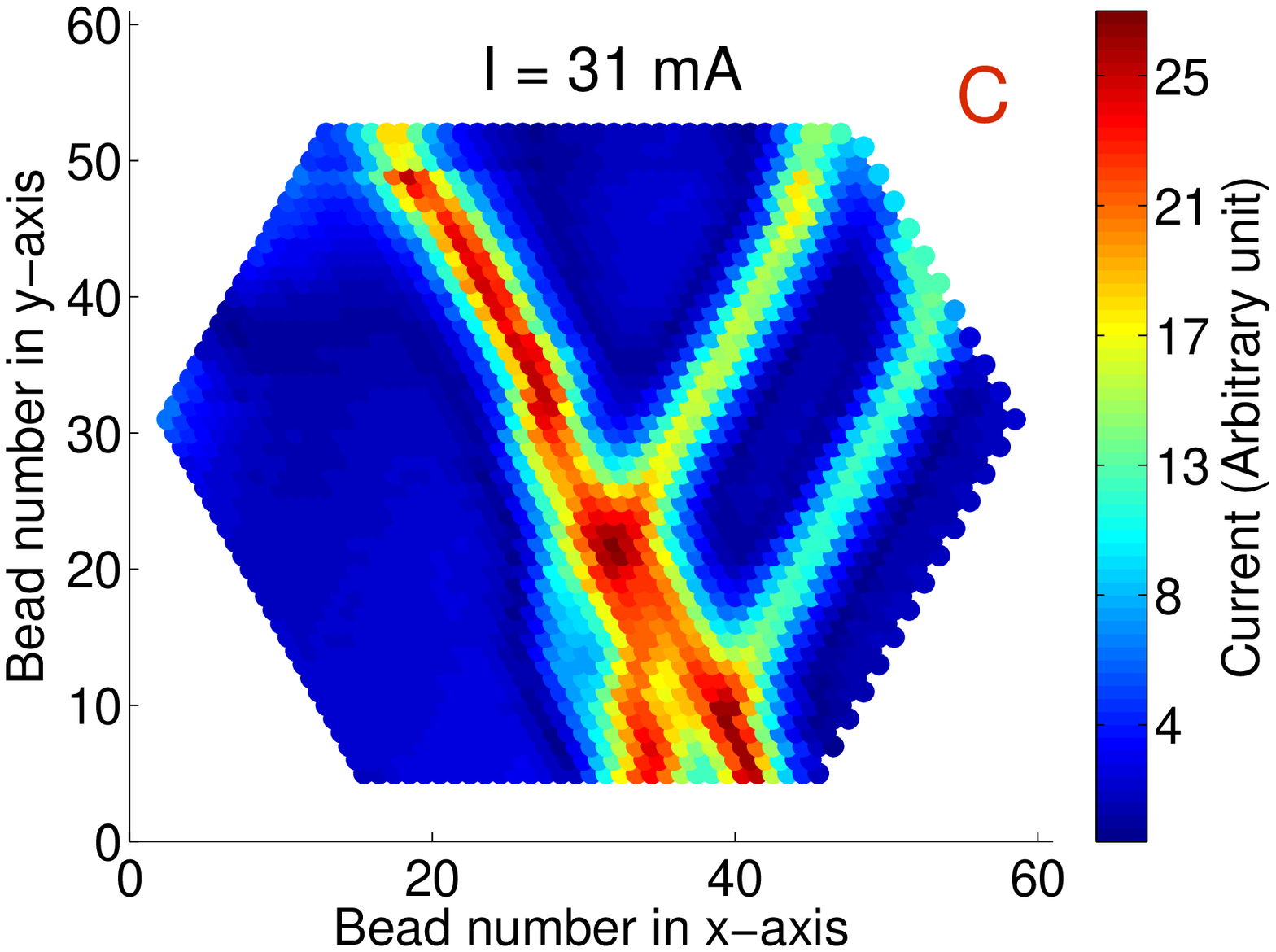}
} \\
\resizebox{.98\columnwidth}{!}{%
 \includegraphics{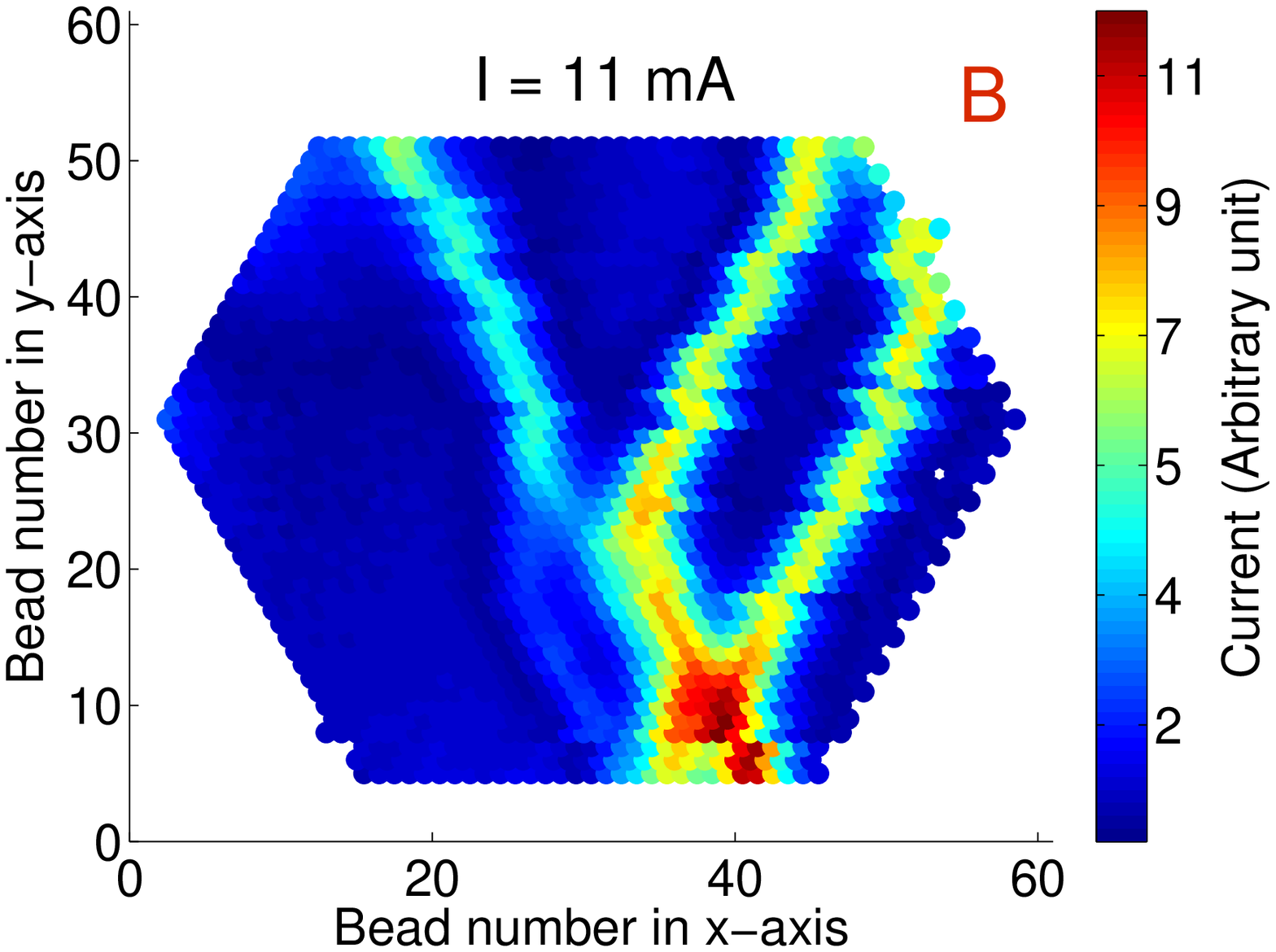}
} \\
\resizebox{.98\columnwidth}{!}{%
 \includegraphics{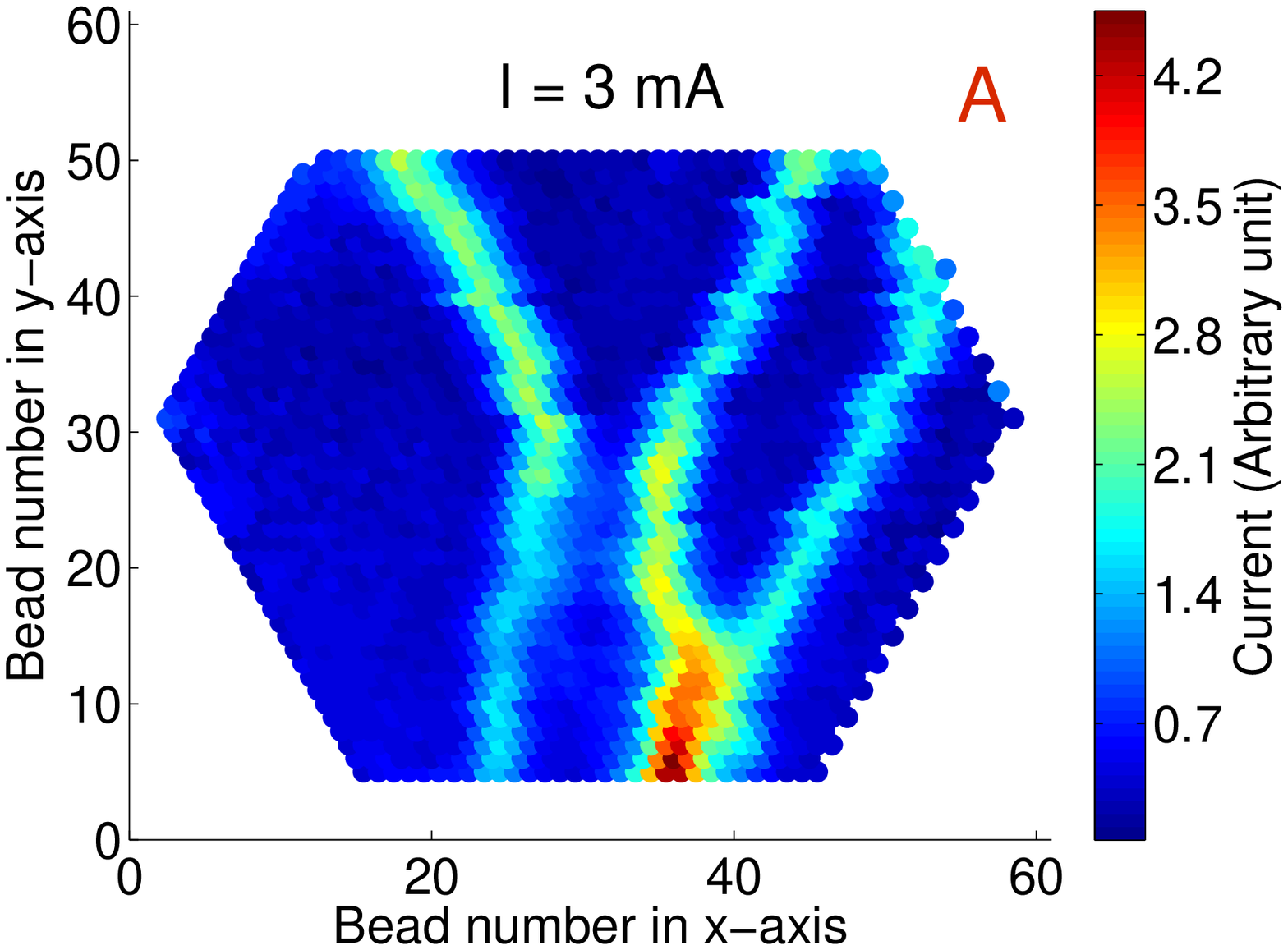}
}
\end{tabular}
\caption{(color online) Visualization of the current path network in an 2D hexagonal packing of metallic beads. The applied voltage increases from bottom to top (corresponding to a total current of $I=3$, 11 and 31 mA and denoted by the letters A, B, and C in Fig.\ \ref{fig02}b). An isotropic stress is created by applying a 20 N force on each upper side. The size of the lattice side is 31 beads. The upper part has not been measured (bead number in the $y$-axis $\geq 52$). Colorbar scales are different in the 3 plots.}
\label{fig03}
\end{figure}

Figures.\ \ref{fig03}a, b and c show the evolution of the current maps when the applied voltage increases. These three maps correspond to the letters A, B and C in the $U$--$I$ characteristic diagram of Fig.\ \ref{fig02}b, that is, to the nonlinear and saturation regime. The current mainly appears concentrated in few discrete paths as shown in Fig.~\ref{fig03}a. When the voltage increases, most of these paths are enhanced, whereas few ones disappear (see Fig.~\ref{fig03}b). For higher applied voltage, the previous paths are strongly enhanced, and a bridge between two paths is even created (see Fig.~\ref{fig03}c). For the highest applied voltage, the total current decreases from 31 to 20 mA (during the 1 hour measurement) due to thermal drift, and the stress decreases of roughly 10\%. The typical observation of the current paths is generic to various stresses within our experimental range, and is not reduced to a small range of compression. The effect of the stress strength on the current path evolution will be described in a future work. 

The current maps of Fig.~\ref{fig03} show that the current paths mainly appear concentrated on linear chains. These preferred paths are reminiscent of stress concentrations (the so-called ``force chains'') carrying an external load imposed on granular systems \cite{inhomogstress,Radjai98,Majmudar05}. The spatial distribution of these force chains depends strongly on the type of loading (isotropic or uniaxial compression, or pure shear) \cite{Majmudar05}. Only the isotropic case is described here. Since the contact resistances strongly vary with the applied force (much more than for an assumed elastic or plastic contact, see Fig.~3 of Ref.\ \cite{Falcon04v1D}), this could explain the similarity between force chains and ``electrical chains". However, as shown below, the contact resistances are also widely distributed for a given force. This random distribution of resistances could also lead to the concentration of current on percolating backbones \cite{Ambegaokar}. The chains of maximum current thus would not be those of maximum stress, except maybe at very high loading. However, we did not perform a simultaneous measurement of force and current maps. Since the resistance distribution is broad, we probe more the contact network (the least resistance paths) than the force network.
\section{Frequency dependence of the saturation voltage}
\label{freq}

Let us now examine the AC electrical response of the 1D chain. As far as the saturation voltage is concerned, two extreme cases can be considered: 

\begin{itemize}
\item Either the contact temperature can follow the voltage variations across a contact. Then the 0.4 V per contact corresponds to the peak amplitude of the AC voltage across a contact. For a sine like variation in the voltage, the rms value of the saturation voltage is  $0.4/\sqrt{2} \simeq 0.28$ V per contact. The temperature variations occur at twice the frequency of the voltage. This doubling temperature frequency stems from the dependence of the contact temperature on the square of the voltage (see below).

\item Or the temperature cannot follow the voltage variations due to the specific heat of the material. Then, the 0.4 V is the rms saturation voltage across the contact.
\end{itemize}

To discuss the intermediate regime between these two extreme cases, let us introduce the influence of the specific heat in the thermal budget. We approximate the temperature $T$ and voltage $U$ fields as spherically symmetric around the center of the metallic bridge. Half of the Joule heating goes in each bead. The current flowing from the left to the right bead, and considering the right bead, we define $u$ as the (negative) difference between the potential at distance $r$ and the potential at the center. Far from the center,  $T=T_o$ and $u=-U/2$. At the center  ($r=0$), $T=T(0)$ and $u=0$. The total heat flow at distance $r$ is
\begin{equation}
-\lambda 2\pi r^2 \frac{\partial T}{\partial r}
\end{equation}
where $\lambda$ is the heat conductivity. The total current $I$ is independent of $r$ and is
\begin{equation}
\label{current}
I=-\frac{2\pi r^2}{\rho} \frac{\partial u}{\partial r}
\end{equation}
where $\rho$ is the electrical resistivity. Joule heating within the considered bead at distance less than $r$ is then
\begin{equation}
-uI=\frac{2\pi r^2}{\rho} \frac{u\partial u}{\partial r} \ {\rm .}
\end{equation}
The thermal budget then becomes
\begin{equation}
-\lambda r^2 \frac{\partial T}{\partial r} = \frac {r^2}{\rho} \frac{u\partial u}{\partial r} - \int_0^rC r^{,2}\frac{\partial T(r^,)}{\partial t}dr^,
\label{gg}
\end{equation}
where $C$ is the specific heat of the bead material.
Using the relation $\lambda\rho=LT$ (see Ref. \cite{Falcon05,Falcon04v1D}), where $L \simeq 2.5\,10^{-8}$~V$^2$/K$^2$ is the Lorentz number, Eq.\ (\ref{gg}) can be written as
\begin{equation}
 T\frac{\partial T}{\partial r} = -\frac{u\partial u}{L\partial r} + \frac{T}{\lambda r^2}\int_0^rCr^{,2}\frac{\partial T(r^,)}{\partial t}dr^, \ {\rm .}
\label{bilan}
\end{equation}
Integrating Eq.\ (\ref{bilan}) from $r=0$ to $\infty$ gives
\begin{equation}
T(\ell)^2-T(0)^2=-\frac{u^2}{L}+\int_0^{\ell} \frac{2T}{\lambda r^2}\int_0^rCr^{,2}\frac{\partial T(r^,)}{\partial t}dr^,dr
\label{Kohlloc}
\end{equation}
or, with $T_m \equiv T(0)$,
\begin{equation}
T_m^2-T_o^2=\frac{U^2}{4L}-\int_0^{\infty} \frac{2T}{\lambda r^2}\int_0^rCr^{,2}\frac{\partial T(r^,)}{\partial t}dr^,dr \ {\rm .}
\label{Kohltot}
\end{equation}
This is the modified Kohlrausch equation due to the second term in the right-hand side of Eq.\ (\ref{Kohltot}) taking the temporal dependence into account. For details on the Kohl\-rausch equation at the thermal equilibrium, see Refs.~\cite{Holm00,Greenwood58} for generality, or Refs. \cite{Falcon05,Falcon04v1D} for this problem.
The influence of this last term is examined in the Appendix A for a sinusoidal voltage $U=U_o\cos\omega t$. We then find
\begin{equation}
T_m^2-T_o^2=\frac{U_o^2}{8L}\left[1+\theta(\omega)\cos (2\omega t-\phi(\omega))\right]
\label{Kohlmean}
\end{equation}
where
\begin{equation}
\theta(\omega)=\frac{1}{\sqrt{1+G(\omega)^2}} \mbox{  ;  }G(\omega)= \frac{4a}{5}\sqrt{\frac{2\omega}{\kappa}}
\label{fff}
\end{equation}
$a=\rho I_o/\pi U_o$ being of the order of the radius of the metallic bridge (see Appendix A). $I=I_o\cos\omega t$ is the first (and main) harmonics of the current, $\kappa$ and $\phi$ being introduced in Appendix A.

As explained in \cite{Falcon05,Falcon04v1D}, the maximum value of $T_m^2$, is fixed to $V_o^2/4L$ by the softening of the metallic bridge, with $V_o \simeq 0.4$V. Substituting this variable into Eq.\ (\ref{Kohlmean}) leads to the rms saturation voltage
\begin{equation}
U_{\rm rms}^2\equiv \frac{U_o^2}{2}=\frac{V_o^2}{1+\theta(\omega)} = V_o^2\frac{\sqrt{1+(32a^2\omega/25\kappa)}}{1+\sqrt{1+(32a^2\omega/25\kappa)}} 
\end{equation}
Defining $I_{\rm rms}\equiv I_o/\sqrt{2}$ and $A\equiv 8\rho/(5\pi V_o)$, the AC current--voltage characteristics for one contact then becomes 
\begin{equation}
U_{\rm rms}^2=V_o^2\frac{\sqrt{1+(A^2I_{\rm rms}^2\omega/\kappa)}}{1+ \sqrt{1+(A^2I_{\rm rms}^2\omega/\kappa)}}
\label{ch}
\end{equation}

\begin{figure}[t!] 
\resizebox{1\columnwidth}{!}{
\includegraphics{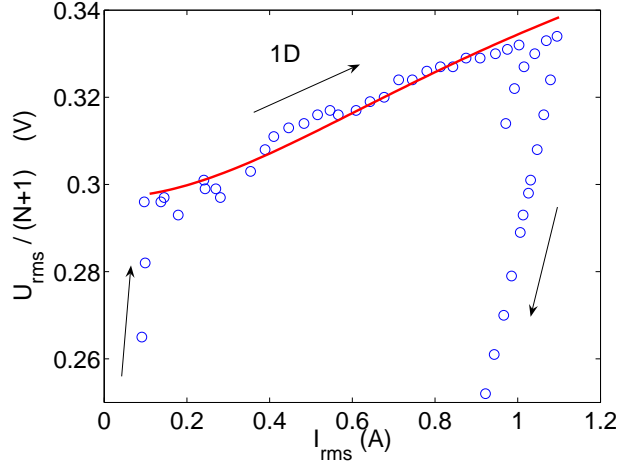}
}
\caption{Comparison between the AC characteristics measured at 1 MHz ($\circ$) in the chain of $N=9$ beads, and the model ($-$) from Eq.\ (\ref{ch}) with $A=1.15$ $\mu$m/A. Only the top part of the experimental curve is shown here to focus on the saturation regime.} 
\label{frequ}  
\end{figure}

Figure \ref{frequ} shows a comparison between measurements at 1MHz ($\circ$-symbols) and Eq.\ (\ref{ch}) (solid line). For this comparison, $\omega=2\pi f$ with $f=1$ MHz, and the value chosen for $A=1.15$ $\mu$m/A is within 30\% of that obtained ($\simeq 0.9$ $\mu$m/A) from the physical properties of the stainless steel beads used in the chain (see Appendix A). $V_o=U_{sat}/(N+1)\simeq0.42$ V is given by the DC saturation measurements. The uncertainties in the measurements do not allow to claim for a full agreement. However, it is clear that the model gives an order-of-magnitude estimate for the frequency effects, and a good explanation of the progressive character of the saturation in such conditions. It also confirms the size of the metallic bridge invoked to explain the saturation plateau: namely $a=\simeq 50$ nm for $I=10^{-1}$ A, and proportional to the current $I$.

\begin{figure}[t!]
\begin{tabular}{c}
\resizebox{1\columnwidth}{!}{%
\includegraphics{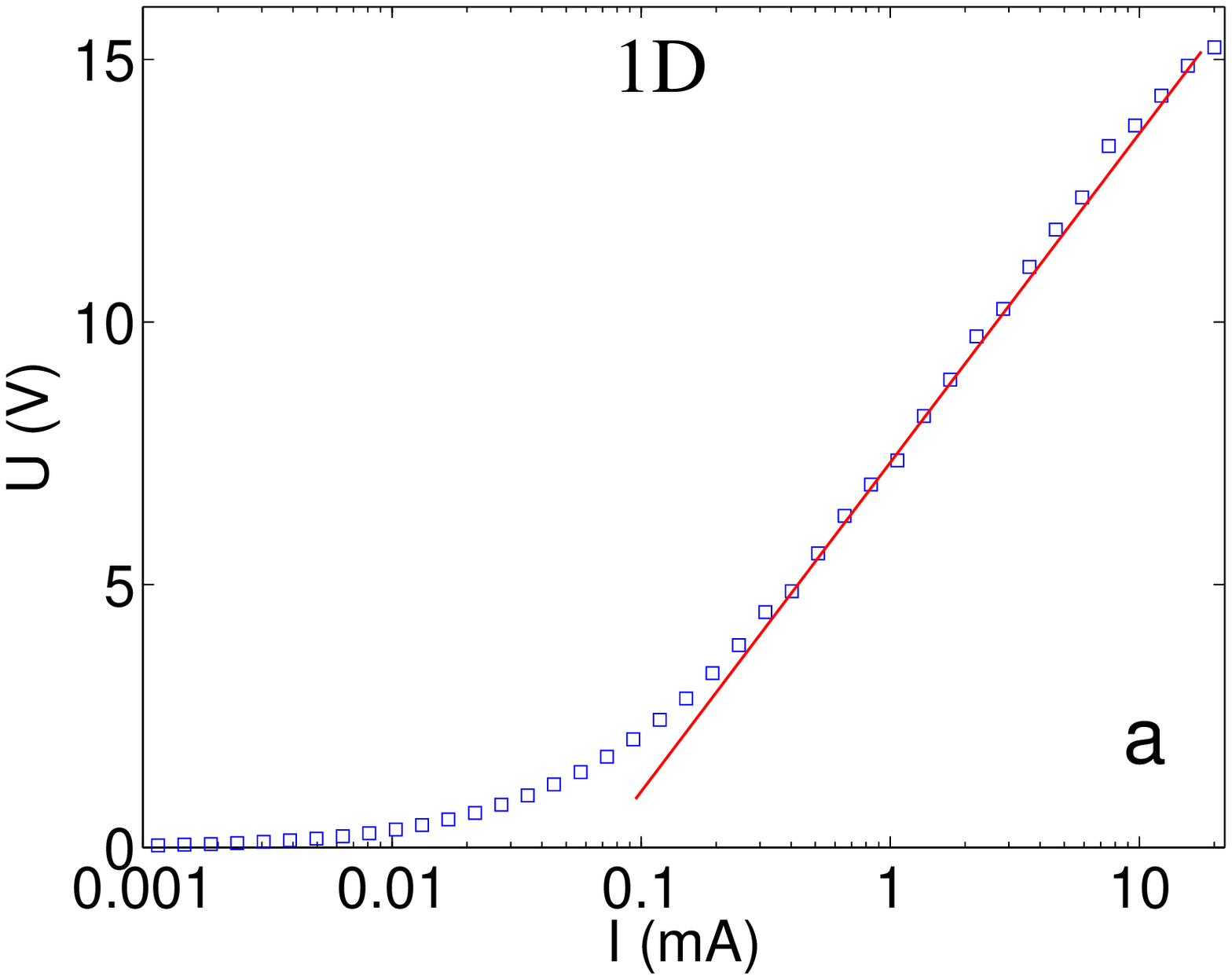}
} \\
\resizebox{1\columnwidth}{!}{%
 \includegraphics{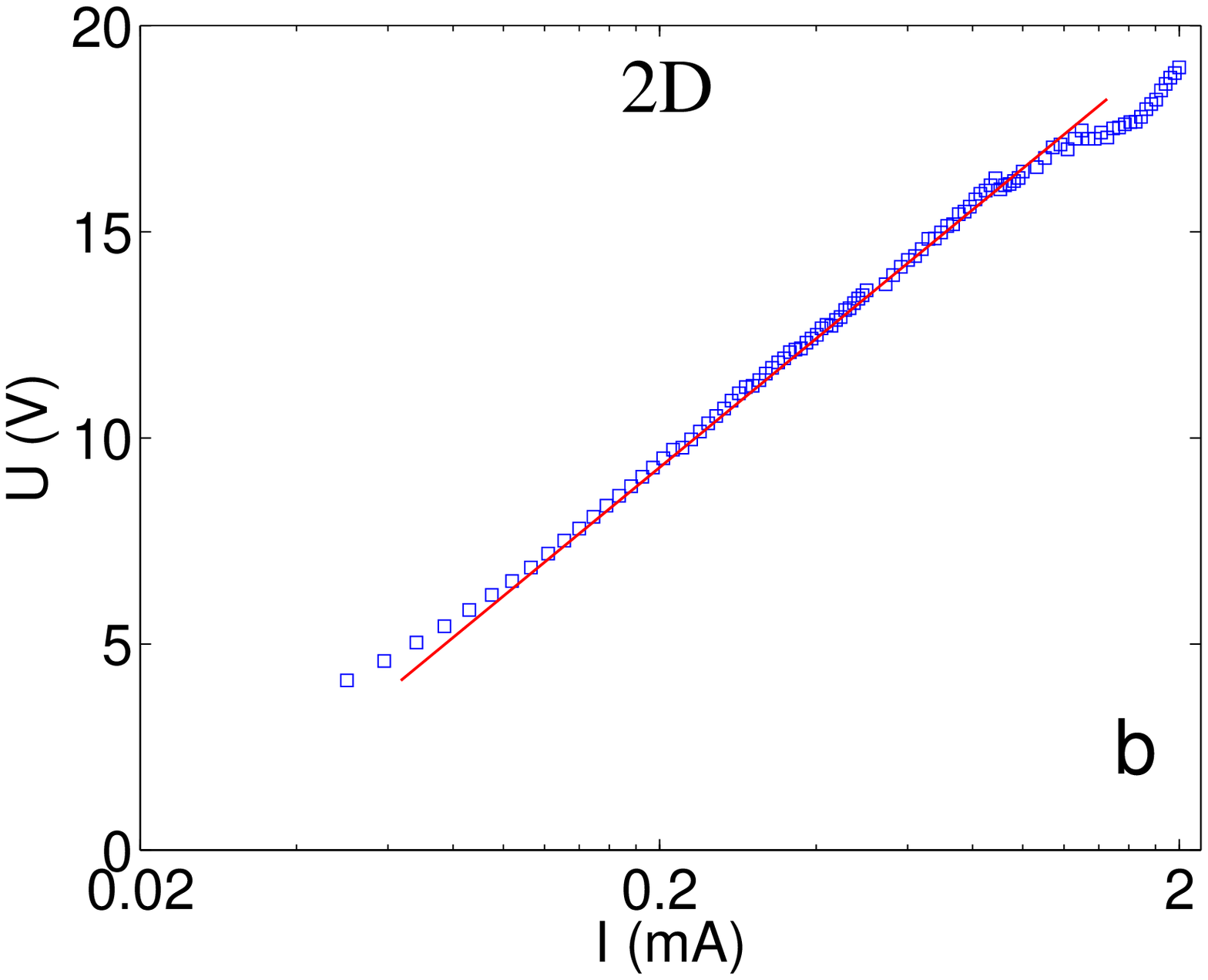}
}
\end{tabular}
\caption{(color online) Zoom of the $U$-$I$ characteristics in Fig.\ \ref{fig02} on a semi-log plot focusing on small currents. {\bf (a)} 1D and {\bf (b)} 2D experiment. ($-$): fits $U \propto \ln{I}$.}
\label{fig05}
\end{figure}

\section{Up-characteristics for $I_N<I<I_S$}
\label{log}

Let us now focus on the intermediate range of the DC applied current $I_N < I < I_S$ (see Fig.\ \ref{schem}). In this case, the voltage across the 1D chain of beads progressively grows up to the saturation voltage. For this current range, the characteristic of the Fig.\ \ref{fig02}a is shown in a semi-log plot in Fig.\ \ref{fig05}a. It can be modeled over roughly two decades as
\begin{equation}
U \propto \ln I \ {\rm .}
\label{ll}
\end{equation}
Only an inhomogeneity of resistances occurs between beads along the chain. In the 2D experiment, there are both stress \cite{inhomogstress,Radjai98,Majmudar05} and resistance inhomogeneities. However, a similar logarithmic characteristic is also observed with the 2D experiment for this range of current, as shown in Fig.~\ref{fig05}b.

We argue that it is simply due to the wide distribution of contact resistances between beads. The rough argument is the following. When the current is progressively raised from zero, the contacts with the largest resistances reach the saturation voltage first. Consequently, a further increase of the current does not result in voltage increase for each of these contacts. On the other hand, the contacts with the lowest resistances do not contribute either to the total rise in voltage. Thus this rise is due to the contacts whose resistance is such that they are close to the saturation voltage $V_o$. This resistance is thus $V_o/I$. For a variation of current $\delta I$, their voltage increase is $\delta U=(V_o/I)\delta I$. By integration, this leads to the log dependence of Eq.\ (\ref{ll}) in qualitative agreement with the observation displayed in Fig.\ \ref{fig05}.

\begin{figure}[h!]
\begin{center}
\resizebox{0.8\columnwidth}{!}{
\includegraphics{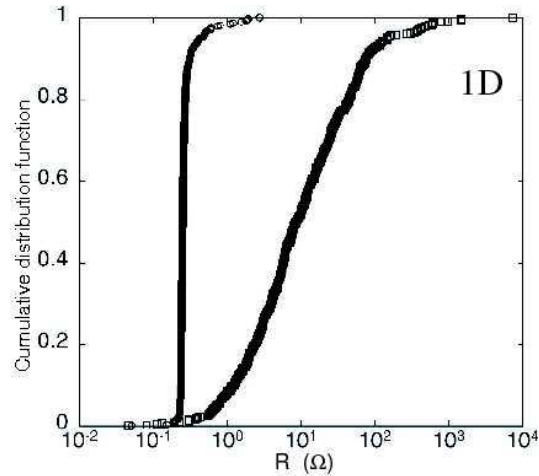}}
\caption{Cumulative distribution function of the resistance of a chain of $N=17$ beads for 480 realizations (in a semi-log plot): before ($\Box$) and after ($\circ$) the saturation regime (see Fig.\ \ref{fig02}a). $\langle R \rangle$ = 38.3 ${\rm \Omega}$ before (0.29 ${\rm \Omega}$ after), and $\sigma_R = $ 154.5 ${\rm \Omega}$ before (0.22 ${\rm \Omega}$ after).}
\label{fig06}
\end{center}
\end{figure}

\begin{figure}[t!]
\begin{center}
\resizebox{0.95\columnwidth}{!}{
\includegraphics{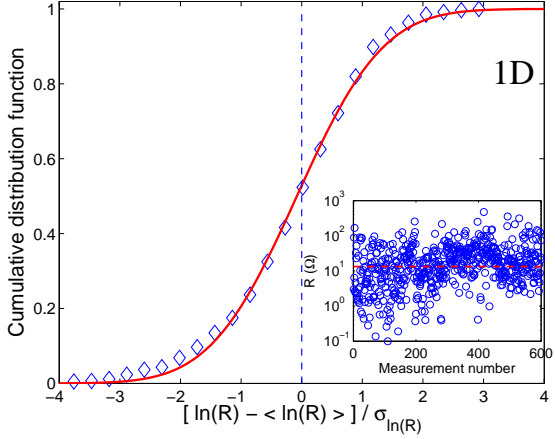}}
\caption{(color online) Inset: values of the resistance, $R$, between 2 beads during 600 realizations before the saturation regime (under stress of 200 N). $\langle R \rangle$ = 28 ${\rm \Omega}$ ; $\sigma_R =$ 43 ${\rm \Omega}$ ; $e^{\langle ln(R) \rangle}$ = 13 ${\rm \Omega}$ ($--$), and $\sigma_{ln(R)} = $1.16. Main: Cumulative distribution function of $ ln(R)$ (centered to the mean and normalized by the rms value). Solid line is a Log-normal fit of mean -0.08 and standard deviation 1.12.}
\label{fig08un}
\end{center}
\end{figure}

A direct measurement of the cumulative distribution of resistances in the chain confirms this broad distribution of contact resistances between beads (see Figs.\ \ref{fig06} and \ref{fig08un}). Figure\ \ref{fig06} shows in a semi-log plot the typical cumulative distribution function of the total resistance of a chain of 17 beads during 480 realizations. The distribution is broad before the saturation regime, and narrower after the saturation. The inset of Fig.\ \ref{fig08un} shows the values of the resistance between 2 beads during 600 realizations before the saturation regime. These values are spread on more than 3 decades. The cumulative distribution function of the logarithms of resistances is well fitted (solid line in Fig.\ \ref{fig08un}) by a log-normal distribution \cite{DaCosta00,Evans}. Even simpler, a flat distribution in resistance logarithms is also a good model, which is equivalent to approximate the cumulative distribution of Fig.\ \ref{fig08un} by its inflexion point tangent.

Such an approximation allows to formalize the above argument. Assume that the resistances $r$ are such that the probability density of their logarithm, $\ell\equiv\ln (r/r_o)$, is $P_o(\ell)=1/(2\ell_{m})$ for $-\ell_{m}<\ell <\ell_{m}$. This uniform distribution $P_o$ is normalized by the constant $\ell_{m}$, and $r_o$ is a constant with the dimension of a resistance. Then, the average voltage per contact is
$$
U=r_oI\int_{-\ell_{m}}^{\ln\left(\frac{V_o}{r_oI}\right)}P_oe^{\ell}d\ell + V_o\int_{\ln\left(\frac{V_o}{r_oI}\right)}^{\ell_{m}} P_od\ell
$$
and
$$
\frac{dU}{dI}=r_o\int_{-\ell_{m}}^{\ln\left(\frac{V_o}{r_oI}\right)}P_oe^{\ell}d\ell = r_oP_o\left(\frac{V_o}{r_oI}-e^{-\ell_{m}}\right)
$$

As $(V_o/r_oI)>>e^{-\ell_{m}}$, we find $dU/dI =P_oV_o/I$ which leads to the expected logarithm characteristic of Eq.\ \ref{ll}. Figure \ref{fig06}  also shows that, once the saturation voltage is reached, all resistances have the same value $V_o/I_{max}$ (see also \cite{Dorbolo07}).

\section{Nonlinear reversible characteristics}
\label{NL}

\begin{figure}[t!] 
\resizebox{1\columnwidth}{!}{
\includegraphics{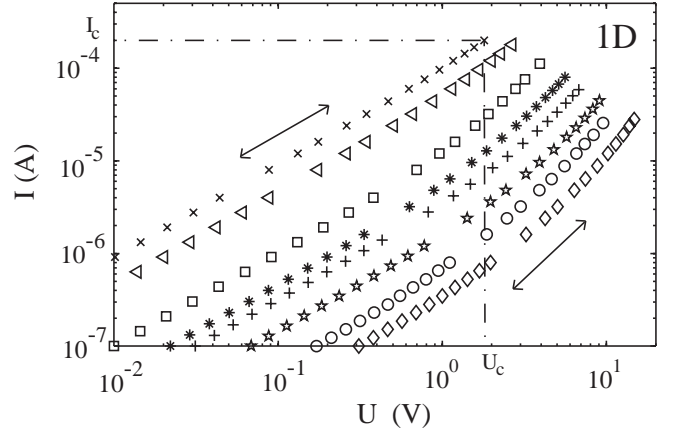}}
\caption{Nonlinear reversible characteristics. The characteristics are reversible up to the point ($U_c$, $I_c$). Different symbols correspond to different applied forces. Arrow means that the process is reversible by decreasing or increasing the current. 1D experiment with $N=40$.} 
\label{fig08}  
\end{figure}

We now examine the part of the characteristics corresponding to $I_L<I<I_N$ (see Fig.\ \ref{schem}). While capturing the main origin of the $\ln I$ behaviour for $I_N<I<I_S$, the model of the previous section is oversimplified. It suggests that the $U$--$I$ characteristics is linear up to the point where the contact with the largest resistance reaches $0.4$V. This is not true. Figure \ref{fig08} shows that non-linearities appear while the $U$--$I$ characteristics remains reversible. 

\begin{figure} 
\resizebox{1\columnwidth}{!}{
\includegraphics{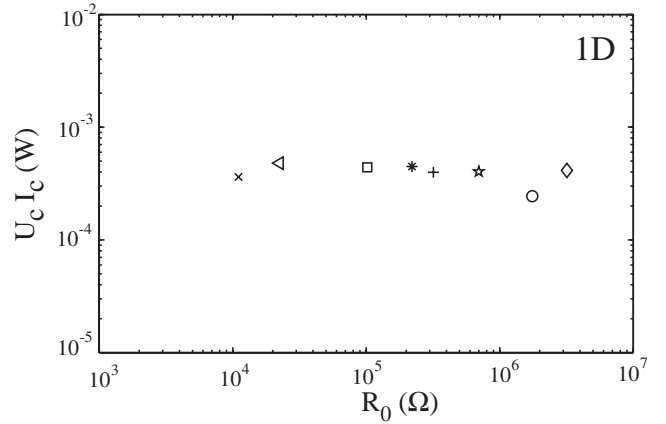}}
\caption{Critical power $U_cI_c$ (for which the characteristics of Fig.\ \ref{fig08} become irreversible) as a function of $R_0$. $R_0$ is the value of the chain resistance at low applied current. 1D experiment with $N=40$. Same symbols as Fig.\ \ref{fig08}.} 
\label{figUcIc}  
\end{figure}

Different applied forces, or even different experiments at the same force, yield different values for the irreversibility threshold (see Fig.\ \ref{fig08}). However, as shown in Fig.\ \ref{figUcIc}, the dissipated power $U_cI_c$ at the irreversibility threshold has approximately the same value $P_c=U_cI_c$ for a very wide range of initial resistances $R_o$, at least for a fixed number of contacts. A similar power-dependent threshold has recently been reported for 3D samples of Copper powder \cite{Falcon04v3D,Creyssels06}. However, here, we do not observe any time evolution of the resistance once the threshold is exceeded, as in the case of 3D samples. While the influence of the dissipated power suggests a thermal origin, we have no precise model for this threshold.

\begin{figure}[t!] 
\resizebox{1\columnwidth}{!}{
\includegraphics{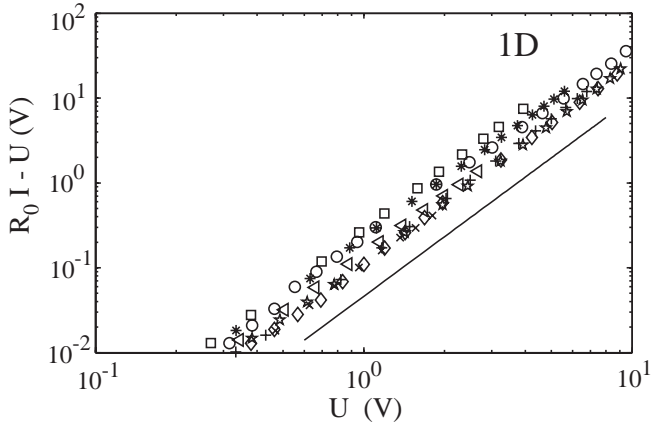}}
\caption{Normalized nonlinear part of the current as a function of the voltage $U$. The logarithmic slope is close to $2.33\simeq7/3$. 1D experiment with $N=40$ beads. Same symbols as Fig.\ \ref{fig08}.} 
\label{fig09}  
\end{figure}

\begin{figure}[t!] 
\resizebox{1\columnwidth}{!}{
\includegraphics{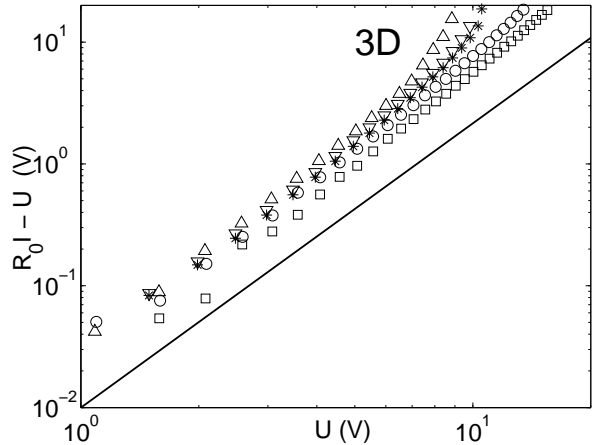}}
\caption{Normalized nonlinear part of the current as a function of the voltage $U$. The logarithmic slope is $7/3$. 3D experiment with Copper powder from Ref.\ \cite{Falcon04v3D}. Symbols corresponds to different forces applied to the powder sample ranging from 640 to 850 N.} 
\label{fig10}  
\end{figure}

In order to characterize the non-linearity, we plot $R_oI-U$ versus $U$ in Fig.\ \ref{fig09}, that is, the nonlinear part of the current, normalized by the initial resistance at low applied current, $R_0$. The logarithmic plot stresses a power-law behaviour of this nonlinear part as
\begin{equation}
R_oI-U \sim U^\alpha
\end{equation} 
The power-law exponent $\alpha$ is always found close to $2.3\simeq 7/3$ whatever the applied force. Let us now use previous data from 3D experiments with Copper powder \cite{Falcon04v3D}, and plot $R_oI-U$ versus $U$ as in Fig.\ \ref{fig10}. Here also we find a similar power-law behaviour of the nonlinear part with roughly the same exponent $\alpha$ than the one found for the 1D experiment of Fig.\ \ref{fig09}.

The situation for 3D Copper powder samples will be clarified in \cite{Creyssels07}. Let us add a few comments about the interplay between the irreversibility threshold occuring at a constant dissipated power ($P_c$), and the saturation threshold ($U=0.4$V per contact). At the former one, $U^2<R_oP_c$, which means that, for sufficiently small initial resistance $R_o$, it will occur before the saturation. However, for large enough $R_o$ (small applied forces), it should occur after. This is impossible, as our explanation of the saturation implies the irreversibility. Indeed, in such situations, we observed that the voltage $U$ can cross the saturation value, and continue to grow until the irreversibility occurs.


\section{Conclusion}
We have reported experiments on the electrical conductivity in 1D and 2D model granular media. The results reported here serve both to confirm previous interpretations and to raise new questions. In both 1D and 2D systems at low current, the wide distribution of contact resistances results in a logarithmic behaviour for the voltage/current characteristics. At high enough current, the dependence of the saturation voltage on the AC frequency confirms the model of the metallic bridge, including the order of magnitude of its size. In the 2D system, both the voltage/current characteristic and the map of the current paths show the quasi-onedimensional nature of the electrical conductivity within a 2D granular medium. The knowledge of the dependence of the current path distribution on different types of loads ({\it e.g.}, pure shear, isotropic compression, ...) will be of primary interest.  Another extension of this work is to understand the nonlinear reversible behaviour of the characteristic observed at very low current in the 1D system. The most remarkable feature is that the nonlinear part of the current is a power law of the voltage with the same exponent that the one found in previous 3D experiments.

\begin{acknowledgement}
A part of this work was supported by the French Ministry of Research under Grant ACI Jeunes Chercheurs 2001. S. Dorbolo is an FNRS  
research associate, and gratefully acknowledges the hospitality of the Physics Laboratory at the ENS Lyon. We thank J.-F. Pinton and H.-C. Nataf for lending us the magnetic field sensor. 
\end{acknowledgement}

\section*{Appendix A}
\label{A}

Our goal is to approximately solve the Kohlrausch modified equation (\ref{Kohltot})
\begin{equation}
T(0)^2-T_o^2=\frac{U^2}{4L}-\int_0^{\infty} \frac{2T}{\lambda r^2}\int_0^rCr^{,2}\frac{\partial T(r^,)}{\partial t}dr^,dr
\label{Kohl}
\end{equation}
with $U=U_o\cos\omega t$.

To wit, we shall use a low frequency approximation for $T(r)$ in the integrals. The imposed voltage across the contact is $U=U_o\cos\omega t$. Thus, $u=-(U_o/2)f(r)\cos\omega t$ and $I=I_o\cos\omega t$. Substituting these two equations into Eq.\ (\ref{current}) gives
$$
I_o=\frac{U_o\pi}{\rho} r^2\frac{\partial f}{\partial r} \ {\rm .}
$$

Neglecting the temperature dependence of $\rho$, we define $a=r^2\partial f/\partial r$, a constant length which is of the order of the radius of the metallic bridge. As $f=1$ for $r=\infty$, the above relation integrates as $f=1-(a/r)$ which is valid for $r>>a$. For $r=0$, $f=0$. We shall thus use the ansatz:
$$
f(r)=1-\frac{a}{r+a} = \frac{r}{r+a}
$$

Let us now limit ourselves to the first harmonics in $2\omega$, that is
\begin{equation}
T(r)=T_{\infty}\left[ h_o(r)+\theta(\omega)h_1(r)\cos(2\omega t-\phi(\omega))\right]
\label{temp}
\end{equation}
with $h_o(0)=1$, $\phi(0)=0$, $\theta(0)=1$ and $\theta(\infty)=0$. 

There is several approximations here. We assume $\phi(\omega)$ is independent of $r$, and $h_1$ independent of $\omega$. However, $\phi(\omega)$ is probably larger when $r$ increases, as the volume of the involved material ($\propto r^3$) is larger. In the same way, $h_1(r)$ is probably smaller at large $r$ for small $\omega$ than for large $\omega$. Both effects have the result to effectively limit the integral to $r$ smaller than the diffusion length $\ell_D$. We thus take them into account by putting the upper limit in $r$ to $\ell_D=\sqrt{\kappa/2\omega}$, with $\kappa\equiv \lambda/C$, $\lambda$ being the thermal conductivity and $C$ the specific heat of the bead material. For stainless steel: $\lambda=16.2$ Wm$^{-1}$K$^{-1}$, $C=3.9\times 10^{6}$ Jm$^{-3}$K$^{-1}$, $\kappa=4.1\times 10^{-6}$  m$^{2}$s$^{-1}$ and $\rho=72\times 10^{-8}$ ${\rm \Omega}$m.

Within the same approximations, at low frequency, Eqs.\ (\ref{Kohlloc}) and (\ref{Kohltot}) then become
$$
T(r)^2-T_o^2=\frac{U_o^2}{8L}(1-f^2)(1+\cos 2\omega t) 
$$
$$
= T_{\infty}^2h_o^2 + 2T_{\infty}^2h_oh_1\cos (2\omega t) - (T_o^2-T_{\infty}^2h_1^2/2)
$$

Neglecting the last term, we obtain
$$
T_{\infty}^2=\frac{U_o^2}{8L} \mbox{ ; } h_o=\sqrt{1-f^2} \mbox{ ; } h_1=h_o/2  \ {\rm .}
$$

Looking only at the $2\omega$ terms, Eq.\ (\ref{Kohl}) then becomes
$$
\theta(\omega)\cos[2\omega t-\phi(\omega)] = \cos (2\omega t) + G(\omega)\theta(\omega)\sin[2\omega t-\phi(\omega)]
$$
with
$$
G(\omega)=4\omega\int_0^{\ell_D} \frac{h_o(r)}{\lambda r^2}\int_0^rCr^{,2}h_1(r^,)dr^,dr \ {\rm ,}
$$
or
\begin{equation}
G(\omega) \simeq \frac{2\omega}{\kappa}\int_0^{\ell_D} \frac{h_o(r)}{r^2}\int_0^rr^{,2}h_o(r^,)dr^,dr   \ {\rm .}
\label{G}
\end{equation}

It thus gives
\begin{equation}
\theta(\omega)=\frac{1}{\sqrt{1+G(\omega)^2}}  \ {\rm .}
\label{theta}
\end{equation}

Keeping only the dominant term in Eq.\ (\ref{G}), thus taking $h_o \simeq \sqrt{2a/r}$, that is
$$
\int_0^rr^{,2}h_o(r^,)dr^, \simeq \frac{2\sqrt{2a}}{5}r^{5/2}\ {\rm ,}
$$
we finally obtain
$$
G \simeq \frac{2\omega}{\kappa}\frac{4a}{5}\ell_D= \frac{4a}{5}\sqrt{\frac{2\omega}{\kappa}}\ {\rm .}
$$

\end{document}